\documentclass[a4paper,10pt,aps,twocolumn,floatfix,unsortedaddress]{revtex4-1}
\usepackage{amsmath,amsfonts,amssymb,graphicx,color,bbm,times}
\newcommand{\be}{\begin{equation}}
\newcommand{\ee}{\end{equation}}
\newcommand{\bea}{\begin{eqnarray}}
\newcommand{\eea}{\end{eqnarray}}

\begin{document}
\title{Quantum diffusion with disorder, noise and interaction}

\author{C. D'Errico$^{1}$}
\author{M. Moratti$^{1}$}
\author{E. Lucioni$^{1}$}
\author{L. Tanzi$^{1}$}
\author{B. Deissler$^{1,2}$}
\author{M. Inguscio$^{1}$}
\author{G. Modugno$^{1}$}
\affiliation{$^{1}$ LENS and Dipartimento di Fisica e Astronomia, Universit\`a di Firenze, and INO-CNR, I-50019 Sesto Fiorentino, Italy}
\affiliation{$^{2}$ Institut f\"{u}r Quantenmaterie, Universit\"{a}t Ulm, Albert-Einstein-Allee 45, D-89069 Ulm, Germany}

\author{M.B. Plenio$^{3}$}
\author{F. Caruso$^{1,3}$}
\affiliation{$^{3}$ Institut f\"{u}r Theoretische Physik, Universit\"{a}t Ulm, Albert-Einstein-Allee 11, D-89069 Ulm, Germany}

\begin{abstract}
Disorder, noise and interaction play a crucial role in the transport properties of real systems, but they are typically hard to control and study both theoretically and experimentally, especially in the quantum case. Here we explore a paradigmatic problem, the diffusion of a wavepacket, by employing ultra-cold atoms in a disordered lattice with controlled noise and tunable interaction. The presence of disorder leads to Anderson localization, while both interaction and noise tend to suppress localization and restore transport, although with completely different mechanisms. When only noise or interaction are present we observe a diffusion dynamics that can be explained by existing microscopic models. When noise and interaction are combined, we observe instead a complex anomalous diffusion. By combining experimental measurements with numerical simulations, we show that such anomalous behavior can be modeled with a generalized diffusion equation, in which the noise- and interaction-induced diffusions enter in an additive manner. Our study reveals also a more complex interplay between the two diffusion mechanisms in regimes of strong interaction or narrowband noise.

\end{abstract}
%\pacs{03.75.Lm, 05.60.-k}

\date{\today}

\maketitle

Disorder, noise and interaction are known to play a fundamental role in the dynamics of quantum systems, but a detailed understanding of their combined action is still missing. The interest in this general problem ranges from electronic systems \cite{kramer}, spin glasses \cite{spin glasses}, and nanoscale quantum Brownian motors \cite{motors}, to quantum communication \cite{PHE2004,CHP2010} and the physics of biological complexes \cite{MRLA2008,PH2008,CLFJ2008,CCDHP2009a,CCDHP2009b}. Despite its importance, there is so far only a very limited theoretical understanding of the interplay of disorder, noise and interaction. Moreover, these ingredients are hard to control in experiments with natural or artificial systems; one exception are quantum optical schemes \cite{White2010,Schreiber2010,Schreiber2012,segev}, where nonlinearities are however weak and do not allow to investigate in depth many-body effects.

The prototypical dynamical problem for disordered quantum systems is the evolution of an initially localized wavepacket \cite{Georges}. To our knowledge this problem has never been studied, neither theoretically nor experimentally, under the combined effect of noise and interaction. It is well known that in one spatial dimension a linear wavepacket is localized in a finite region of space by the Anderson localization mechanism. Noise is instead known to break the coherence that is necessary to achieve localization, giving rise to a diffusive expansion of the wavepacket, as predicted by several theoretical approaches \cite{ovchinnikov75,Madhukar77,ott,shepelyansky,cohen} and also observed in experiments with atoms and photons \cite{bayfield,blumel,walther,Steck,photons}. Finally, also a weak interaction can inhibit the Anderson localization, through the coherent coupling of single-particle localized states \cite{Shepe93,Kopidakis08,Pikovsky08,Flach09,LDM2009,flach,kolovsky,wellens,finkelstein,cherroret,deissler2010}, giving rise to a subdiffusion, i.e. a time-dependent diffusion coefficient, that has recently been observed in experiments with ultracold atoms \cite{lucioni10}. A stronger interaction, in systems with limited kinetic energy such as lattices, can instead lead to other localization phenomena, such as self trapping \cite{smerzi97} or the Mott insulator \cite{mott68}. In this regime the presence of disorder can give rise to new quantum phases \cite{giamarchi87,fisher89}. However, while both microscopic theories and macroscopic models exist for noisy or many-body disordered systems, no such theory has been developed for the evolution of a wavepacket in disorder with the simultaneous presence of noise and interaction effects. Even an intuitive understanding of the problem is prevented by the difficulty in combining the incoherent dynamics generated by noise with the coherent coupling due to interaction.

In this work we employ an ultracold Bose-Einstein condensate in a disordered optical lattice to investigate the general features of the expansion of a wavepacket subjected to controlled broadband noise and weak repulsive or attractive interaction. In general, we find that the combination of noise and interaction gives rise to a faster, anomalous diffusion of the wavepacket, with a time-exponent that depends in a complex manner on the system parameters. Our extensive exploration, supported by numerical simulations, indicates that such dynamics can be modeled quite accurately over a wide range of parameters by a generalized diffusion equation in which the two diffusion terms due to noise and interaction simply add up. This surprisingly simple result seem to persists also in regimes where a perturbative modeling of the individual diffusion mechanisms breaks down. For particularly strong interaction strengths or narrowband noise, we find instead a more complex interplay of noise and interaction in the expansion, which cannot be modeled in a simple way and will require a more detailed investigation. Besides providing a general model for the diffusion in an interacting noisy system, this study highlights the capability of quantum gases in disordered optical potentials to investigate other general problems related to noise in quantum systems.\\

\begin{large}\textbf{Results}\end{large}

\begin{figure}[t]
\includegraphics[width=1\columnwidth]{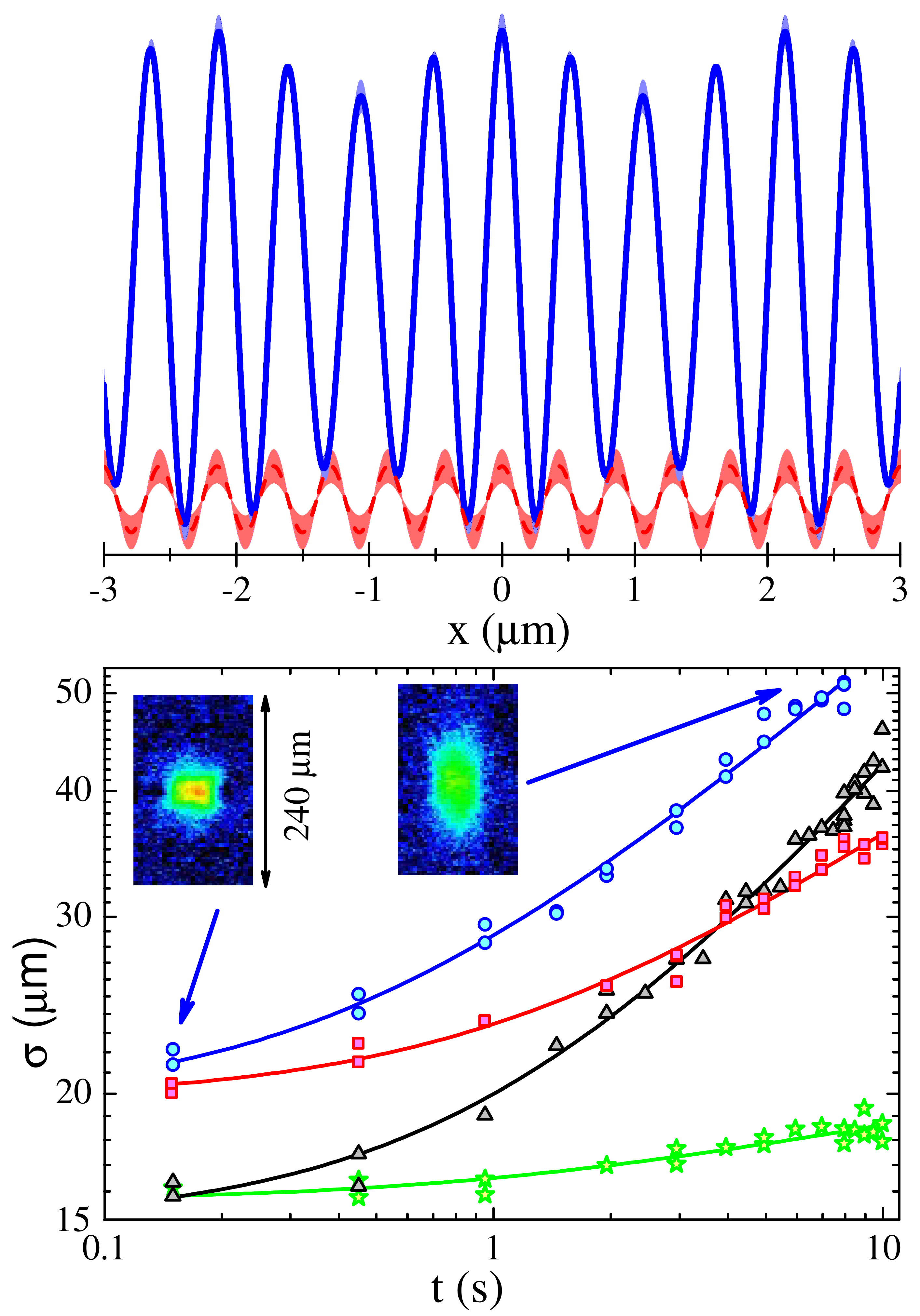}
\caption{\textbf{Expansion with disorder, noise and interaction.} a) Noisy quasiperiodic potential (solid line), realized by a static main optical lattice overlapped to a secondary one (dashed line), with a broadband amplitude modulation. b) Measured time evolution of the width for different values of noise amplitude and interaction energy for $\Delta/J=4$: $A=0$ and $E_{int} \sim 0$ (stars), $A=0.9$ and $E_{int} \sim 0$ (triangles), $A=0$ and $E_{int} \sim 0.8 J$ (squares) and $A=0.8$ and $E_{int} \sim 0.8 J$ (circles). The lines are fits with Eq. (\ref{eq.fit}).} \label{fig1}
\end{figure}

\textbf{Dynamics in a disordered lattice.}
The experiment is based on a Bose-Einstein condensate of $^{39}$K atoms in the fundamental energy band of a quasiperiodic potential, which is generated by perturbing a strong primary optical lattice with a weak, incommensurate, secondary one (Fig.~\ref{fig1}a). The site-to-site tunneling energy $J$ of the main lattice and the disorder strength $\Delta$ characterize the corresponding Hamiltonian \cite{roati2008}. In the non-interacting case the system shows an Anderson-like localization transition for a finite value of the disorder $\Delta=2J$. Above this threshold all eigenstates are exponentially localized, with a localization length $\xi\approx d/\ln(\Delta/2J)$ \cite{Aubry}, where $d$ is the spacing of the main lattice.

The noise is introduced by an amplitude modulation of the secondary lattice, with controllable strength $A$ and with a frequency $\omega_m$ that is randomly varied in a proper interval (Methods). This corresponds to a broadband spectrum with a controllable width of the same order of the energy bandwidth  of the quasiperiodic potential, i.e. $W \approx 2 \Delta + 4 J$. We note that our noise is non-dissipative, and in general we have an out-of equilibrium situation in which the fluctuation-dissipation relation does not hold. However, the finite bandwidth of the lattice sets a limit of the order of $4J$ to the maximum kinetic energy that can be pumped into the system by the noise source. Numerical simulations we performed indicate that after a typical time $t=0.1-1$s our system has reached this limit, and most of the long-time dynamics happens in a quasi-equilibrium regime. \\

A magnetic Feshbach resonance allows to control the contact interaction \cite{Roati07}, by tuning the s-wave scattering length $a$ and in turn the mean interaction energy per particle $E_{int}\approx2\pi\hbar^2 a \bar{n}/m$, where $\bar{n}$ is the mean density and $m$ is the particles mass.

To study the dynamics, we initially prepare the condensate close to the ground state of the lattice, also in presence of a tight axial trap. We then switch off the trap and study the expansion dynamics along the lattice, thanks to an additional radial confinement. To do so, we detect the axial density profile $n(x,t)$ with destructive absorption imaging at increasing times $t$, up to $t=10$~s. The measurements we present are typically referred to as the root mean square width $\sigma$ of the sample along the lattice.\\

\textbf{Observation of anomalous diffusion induced by noise and interaction.}

A typical time evolution of the width of the system for $\Delta>2J$ in presence of noise and interaction is shown in Fig.~\ref{fig1}. The disorder-induced localization is broken by interaction or noise alone and also by their combination. In all cases, we observe a short-time transient that evolves into an asymptotic behavior, which is different in the three cases. To analyse the expansion we start by recalling that a solution of the diffusion equation
\be \label{eq.diff}
\frac{\partial n(x,t)}{\partial t}=\frac{1}{2}\frac{\partial}{\partial x}\Big{(}D\frac{\partial n(x,t)}{\partial x}\Big{)}\,,
\ee
is a Gaussian distribution $n(x,t)\approx\exp{[-x^2/2\sigma(t)^2]}$, with $d\sigma^2(t)/dt=D$.
This would imply a time dependence of the distribution width as
\be \label{eq.fit}
\sigma (t) = \sigma_{0} (1+t/t_{0})^{\alpha}\,,
\ee
with the time-exponent $\alpha$=0.5. For our experimental distributions, we find that a solution of a generalized diffusion equation of the form of Eq.~(\ref{eq.fit}) with $\alpha$ as a free parameter, can be used to fit the rms width of $n(x,t)$ in all cases. Here $\sigma_0$ is the initial width, $t_0$ is a free parameter that represents the crossover time from the short-time dynamics to the asymptotic regime and the exponent $\alpha$ characterizes the expansion at long times. In our data we typically see around one decade (in time) of asymptotic expansion.
In presence of noise alone we typically observe normal diffusion, i.e. $\alpha=0.5$, as expected in case of white noise. The dynamics is instead subdiffusive, i.e. $\alpha<0.5$, in presence of a repulsive interaction; as we will discuss later, this is essentially due to a reduction of the interaction coupling as the system expands. In presence of both noise and interaction we typically observe a non trivial anomalous diffusion, with expansion exponent $\alpha$ depending on the relative value of interaction energy $E_{int}$ and noise amplitude $A$. As we discussed above, we are not aware of any theory for this combined problem. We therefore model the expansion with a generalized diffusion equation for $\sigma^2 (t)$ where the instantaneous diffusion coefficient is the sum of the two coefficients of interaction and noise alone, i.e.
\be \label{eq.diff_sum}
\frac{d \sigma^2 (t)}{d t} = D_{noise} + D_{int}(t)\,.
\ee
Here $D_{noise}$=const is the diffusion coefficient due to noise alone, while $D_{int}(t)$ is the time-dependent diffusion coefficient for interaction alone. A main result of this work is that such general diffusion equation is valid in a wide range of parameters, as we will show in detail in the following through an analysis of the individual diffusion mechanisms and their combination. A similar generalized diffusion equation has been theoretically predicted for Brownian motion of classical interacting particles \cite{brow}.\\

\textbf{Noise-induced normal diffusion.}
Let us start by exploring in detail the effect of noise on the dynamics of the linear system, which we realize by tuning the scattering length $a$ close to zero. As shown in Fig.~\ref{fig1}b, we observe that a finite noise amplitude $A\neq0$ results in a slow expansion of the initially localized sample. The shape of $n(x)$ keeps being Gaussian at all times. We fit $\sigma(t)$ with Eq.~(\ref{eq.fit}); from these and other data we measure $\alpha=0.45(5)$, which is therefore consistent with normal diffusion, and we extract a diffusion coefficient $D=\sigma_0^2/t_0$.

\begin{figure}[t]
\includegraphics[width=1\columnwidth]{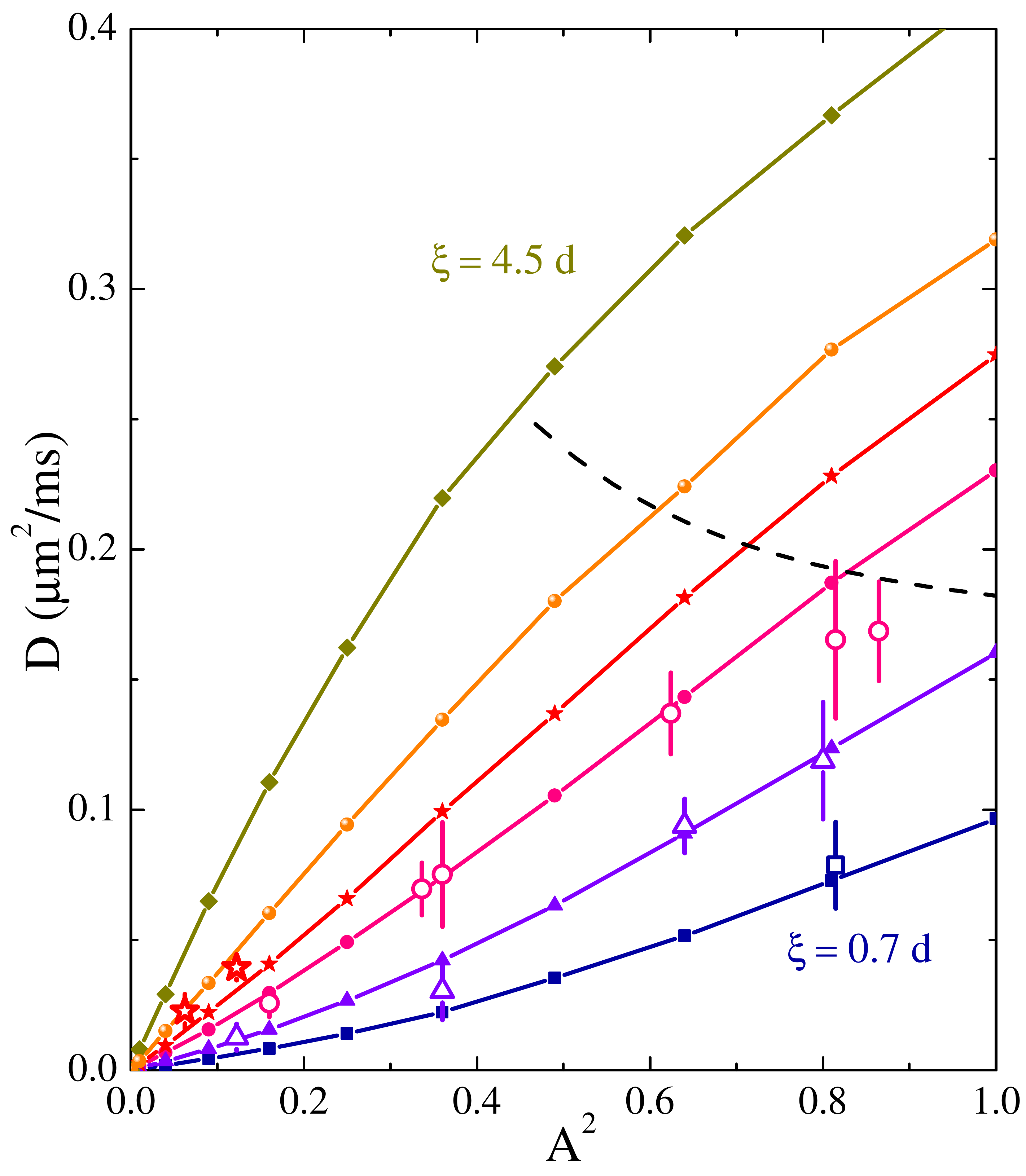}
\caption{\textbf{Noise-induced normal diffusion: $\xi$ and $A$ dependence of the diffusion coefficients.} a) Measured diffusion coefficient $D$ vs the square of the noise amplitude $A$, for different values of the disorder strength. From top to bottom: $\Delta/J=2.5$ (rhombuses), $3$ (spheres), $3.5$ (stars), $4$ (circles), $5.5$ (triangles) and $8$ (squares). The open symbols stand for the experimental measurements and the filled ones for numerical simulations. The dashed line represents the critical values for the noise strength $A_c$ and $D$ above which the perturbative approach is no longer valid (Methods).}  \label{fig2}
\end{figure}

We have performed an extensive investigation of the diffusion dynamics, by exploring different values of the noise amplitude $A$ and the disorder strength $\Delta/J$, i.e. different localization lengths. The measured diffusion coefficients $D$, shown in Fig.~\ref{fig2}, are in good agreement with numerical simulations in terms of a generalized Aubry-Andr\'e model \cite{Modugno09,LDM2009}, including a dynamical disorder analogous to the experimental one (Methods).
We can interpret the observed diffusion as an incoherent hopping between localized states driven by the broadband noise. A perturbative approach suggests $D\propto \Gamma \xi^2$ and $\Gamma \propto A^2$, where the localization length $\xi$ represents the natural length-scale of the hopping and $\Gamma$ is the perturbative transition rate \cite{shepelyansky}. For the Aubry-Andr\'e model, in the limit of a noise bandwidth equal to the lattice bandwidth, we calculate a diffusion coefficient:
\be \label{eq.pertDCoeff}
D \propto \frac{A^{2} J}{\hbar} \frac{(\xi+d)^{2}}{1+e^{d/\xi}}\; ,
\ee
which is in good agreement with both the experiment and numerical simulations (Methods).
As shown in Fig.~\ref{fig2}, we observe the linear dependence of $D$ on $A^2$ and also an increase of $D$ with $\xi$.

The described behavior persists in most of the range of values of $A$ and $\xi$ we have been able to explore in the experiment. The simulations however give a clear indication that large values of $A$ and/or $\xi$ would bring the system into a different regime, where the dependence on the localization length becomes weaker. This is expected since, in the presence of a sufficiently strong noise, the perturbative approach based on localized states must fail. Interestingly, the experimental data, the numerical ones and the perturbative model indicate that the crossover to this second regime happens at strong noise amplitudes, $A\approx 1$, for our range of rather short localization lengths, i.e. $\xi \approx d$.\\

\begin{figure*}[t]
\includegraphics[width=1\textwidth]{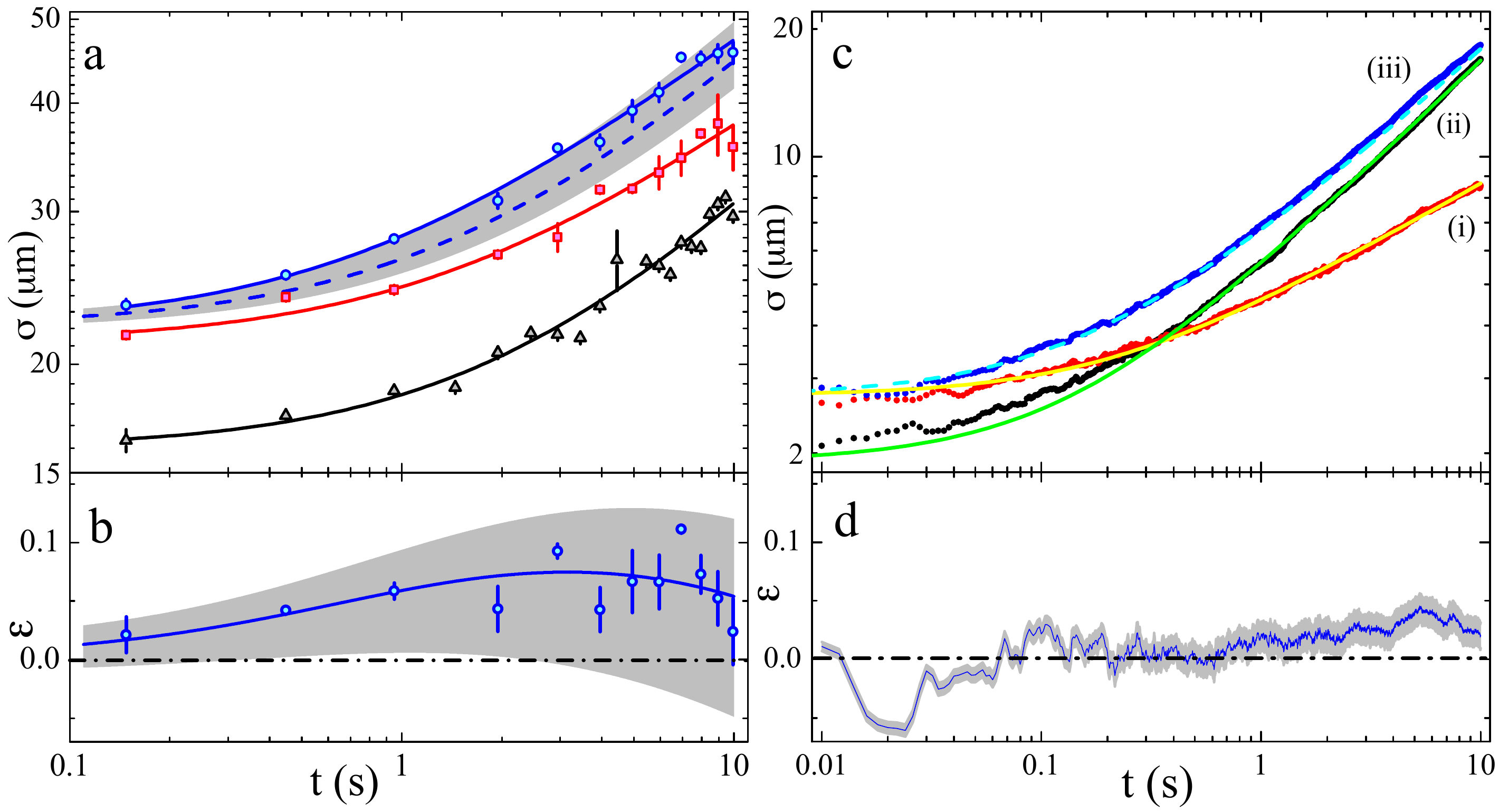}
\caption{\textbf{Noise and interaction additive anomalous diffusion.} a) Measured time evolution of the width for noise alone (triangles), interaction alone (squares), or both (circles), for $\Delta/J=4$. The noise amplitude is $A=0.6$, while the interaction strength is $E_{int} \sim 0.8 J$. The experimental data are fitted with Eq. (\ref{eq.fit}) (solid lines). The dashed line is the numerical solution of Eq. (\ref{eq.diff_sum}) using the extracted diffusion coefficients from the fits, with the confidence interval shown as a grey area.  b) Relative residuals of the fit, for the case with noise and interaction, and the solution of Eq. (\ref{eq.diff_sum}), with the associated confidence interval (grey area).  c) Numerical simulation of the time evolution of the width for interaction alone (i), noise alone (ii), and both (iii), for $\Delta/J=2.5$, $J = 180$~Hz, $A=0.195$, $T=1$ ms and $E_{int}=1.6J$. The numerical data are fitted with Eq. (\ref{eq.fit}) (solid lines), whereas the dashed line is the numerical solution of Eq. (\ref{eq.diff_sum}).  d) Relative residuals of the numerical $\sigma(t)$ with noise and interaction (iii) and the solution of Eq. (\ref{eq.diff_sum}). The grey area represents the confidence level of 2 standard deviations.}  \label{fig3}
\end{figure*}

\textbf{Interaction-induced subdiffusion.}
We now discuss the effect of the interaction, which is introduced in our system by changing the scattering length $a$ to a finite value, hence introducing a finite $E_{int}$. The effect of such non-linearity on the dynamics on a lattice with static disorder has already been studied in theory \cite{Shepe93,Kopidakis08,Pikovsky08,Flach09,LDM2009,flach,kolovsky} and experiments \cite{photons,lucioni10}. Basically, the finite interaction energy breaks the orthogonality of the single-particle localized states, weakening the localization. In this case one can describe perturbatively the resulting dynamics as an  interaction-assisted coherent hopping between localized states, with a coupling strength that decreases as the sample expands, since the density and hence $E_{int}$  decrease \cite{lucioni10}. This results in a subdiffusive behavior, i.e. in a diffusion with a decreasing instantaneous diffusion coefficient
\begin{equation}
D_{int}(t)=2\alpha_{int} \frac{\sigma_0^{\alpha_{int}^{-1}}}{t_0} \sigma(t)^{2-\alpha_{int}^{-1}}\,,
\end{equation}
where $\alpha_{int}$ is the time-exponent. From $d\sigma(t)^2/dt=D_{int}(t)$ one gets indeed an evolution of the form $\sigma(t)=\sigma_0(1+t/t_0)^{\alpha_{int}}$. The precise value of such exponent depends on the details of the interaction strength and of the spatial correlations of the disorder: for an uncorrelated random disorder and $E_{int}\approx\Delta$ a perturbative approach predicts $\alpha_{int}=0.25$. In our experiment we find exponents in the range 0.2$<\alpha_{int}<$0.35. A detailed description of the various regimes achievable in a quasiperiodic lattice can be found in Ref. \cite{larcher}.  \\

\textbf{Generalized diffusion model.}
When we introduce noise and interaction at the same time, we observe an expansion that is globally faster than for noise or interaction alone, but has an exponent $0.3<\alpha'<0.5$ which is intermediate between the two previous ones. This indicates that both delocalization mechanisms are playing a role in the expansion. One example over many of these observations is shown in Fig.~\ref{fig3}, which reports three characteristic expansion curves for noise or interaction alone, or both. In presence of both noise and interaction, we also find that the distribution does not manifestly deviate from a Gaussian during the expansion. To analyse the combined dynamics we use the generalized diffusion equation of Eq.~(\ref{eq.diff_sum}), where $D_{noise}$ and $D_{int}(t)$ are separately extracted by fitting respectively the case with noise alone and interaction alone with Eq.~(\ref{eq.fit}). The numerical solution of the differential equation Eq.~(\ref{eq.diff_sum}) is actually in good agreement with the experimental data and also with the numerical results of the theoretical model, as shown in Fig.~\ref{fig3}.

These examples of the experimental observations and numerical simulations strongly support the hypothesis of additivity of the two delocalization mechanisms, at least at first order. In the experimental data (see Fig.~\ref{fig3}a-b) we typically observe a slightly faster diffusion with noise and interaction. This difference is due to the axial excitation in presence of noise, which in turn excites the radial degrees of freedom in presence of interaction \cite{lucioni10} and produces the increased expansion we observe in Fig.~\ref{fig3}b (see Supplementary Information for details). On the contrary, in numerical simulations, which neglect the axial-to-radial coupling, there are much smaller deviations from the solution of Eq.~(\ref{eq.diff_sum}), of the order of a few lattice constants (see Fig.~\ref{fig3}d).

The good agreement between the experimental data and the prediction of Eq.~(\ref{eq.diff_sum}) persists for the whole range of $A$ and $E_{int}$ that are accessible in the experiment, with $A$ ranging from 0.4 to 1 and $E_{int}$ adjustable up to $\approx J$. Furthermore, as we described above (Fig.~\ref{fig2}), in numerical simulations we can explore the region of large $A$ and/or $\xi$, where the perturbative description of the noise effect fails. Interestingly enough, although this regime is not perturbative, we find the additivity of noise and interaction mechanisms also in this case.

An investigation over a broad range of parameters in theory indicates that the system should behave in a fully symmetric way for attractive and repulsive interaction. We have experimentally tested this expectation with a sample prepared with attractive interaction ($a\approx$-100$a_0$ with $a_0$ being the Bohr radius), where the behavior is fully analogous to the one in Fig.~\ref{fig3}a-b, i.e. the two sources of delocalization simply add up. Note that systems prepared with $a<0$ have not a stable state at low energy, but typically occupy the whole energy band from the beginning of the expansion. This however does not seem to affect the expansion dynamics.\\

Therefore, the large region of validity of Eq.~(\ref{eq.diff_sum}) supports the idea that the observed anomalous diffusion is driven by the additivity of the two delocalization mechanisms, and indicates that the incoherent noise-induced hopping between localized states does not destroy their coherent coupling due to the interaction. Note that the interaction-assisted diffusion tends to vanish as the sample expands, so that one should expect a long-time crossover to a regime where interaction effects are negligible, and the system diffuses normally due to noise alone. While we can clearly observe this crossover in the simulations, its experimental characterization is prevented by the limited observation time.\\

\textbf{Regimes of noise-interaction interplay.}

\begin{figure}[t]
\includegraphics[width=1.\columnwidth]{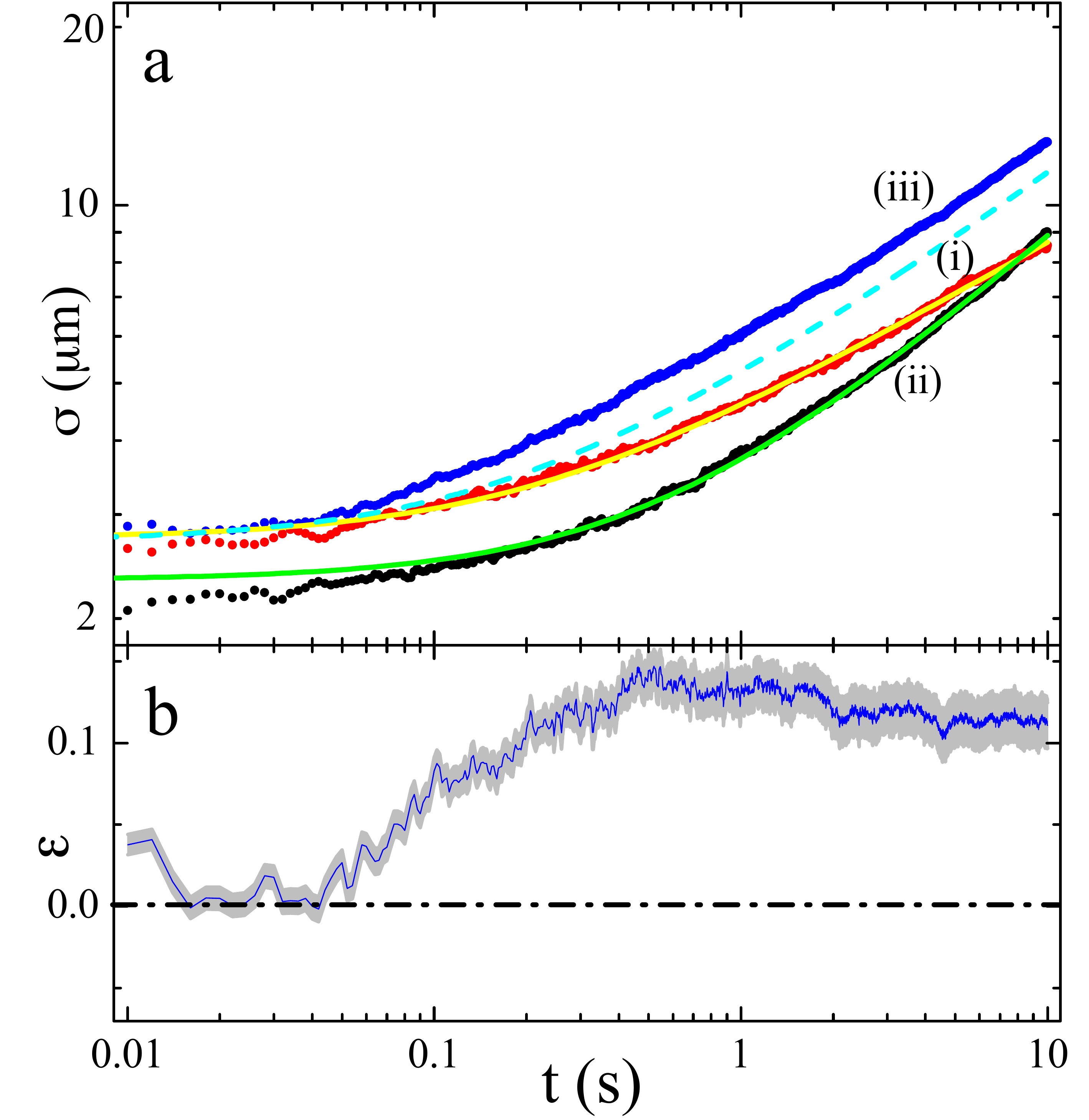}
\caption{\textbf{Deviations from additivity with a narrowband noise.} Numerical simulation of the time evolution of the width (a) and relative residuals (b) with the same parameters of Fig. $\ref{fig3}$c-d, but with $T=10$ ms. Deviations from additivity are evident in the discrepancy between the curve with noise and interaction (iii) and the solution of Eq. (\ref{eq.diff_sum}) (dashed line).}  \label{fig4}
\end{figure}

\begin{figure}[t]
\includegraphics[width=1.\columnwidth,clip]{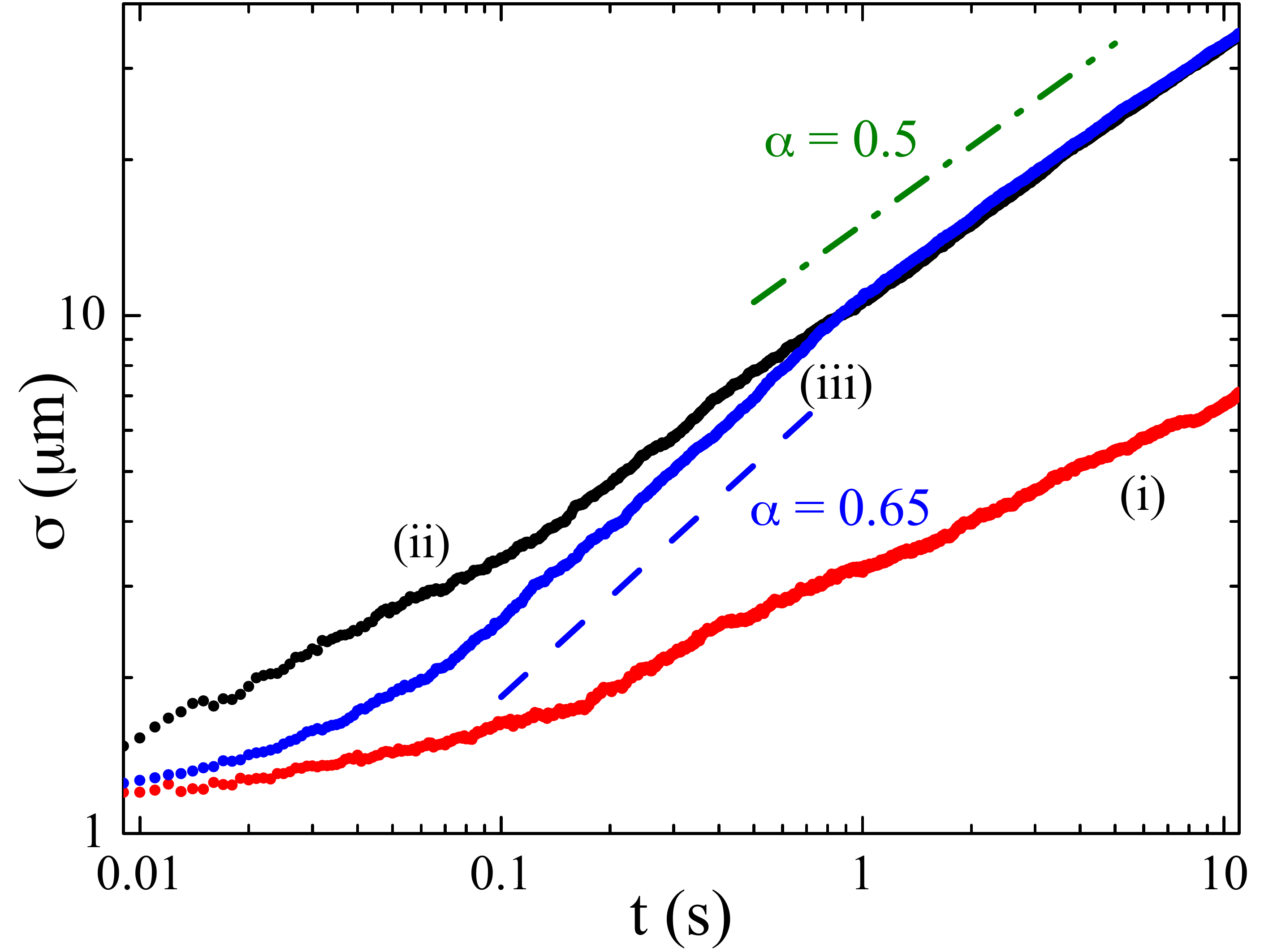}
\caption{\textbf{Deviations from additivity in the self-trapping regime.} Numerical simulations of the time evolution of the width for attractive interaction alone (i), noise alone (ii), or both (iii), for $\Delta/J=2.5$, $J=180 Hz$, $A=0.4$, $\beta=-35$, $T=0.6$ ms. An intermediate super-diffusive ($\alpha>0.5$) regime seems to be induced by self-trapping, while asymptotically one has normal diffusion with exponent $\alpha\sim 0.5$. The (green) dashed-dotted line shows the diffusive slope, while the superdiffusive one is represented by the (blue) dashed line.} \label{fig5}
\end{figure}

While all experimental observations are well described by the generalized diffusion model discussed above, our numerical simulations indicate that deviations might appear when the system is subjected to a trapping mechanism. This can happen in both cases of noise with a bandwidth $\delta \nu$ much smaller than the system one $W/h$, or  of an interaction that is strong enough to produce self-trapping \cite{smerzi97,kolovsky,LDM2009}. In both cases, the simulations show an interplay of noise and interaction which cannot be described by Eq. (\ref{eq.diff_sum}).

In the first regime, i.e. small noise bandwidth ($\delta \nu \ll W/h$), the diffusion is much slower than for a broad noise spectrum ($\delta \nu \approx W/h$), because just a small fraction of the localized states can be coupled by the noise. Very interestingly, the addition of the interaction in this regime seems to restore a fast expansion, as if the additional coupling mechanism between localized states provided by the interaction would effectively broaden the noise spectrum. One example of this behaviour is shown in Fig.~\ref{fig4}, where $\delta \nu$ is $10$ times smaller than that in Fig.~\ref{fig3}c-d. One observes a noise-induced diffusion coefficient $D$ that is smaller than in the case of broadband noise. Interaction is able to restore a diffusion comparable with the one obtained with interaction and broadband noise, therefore much faster than the one predicted by the additivity conjecture.

Note that in this regime the additivity hypothesis fails despite of the validity of our perturbative approach to describe the noise-induced diffusion. This example and the complementary case of strong noise which manifests additivity to the interaction, although the perturbative model is not valid, indicate an independence of the additivity conjecture from the validity of the perturbative description of the noise.

The second regime, i.e. self-trapping, is a general scenario for the expansion in lattices in presence of interaction, which is reached when $E_{int}$ is too large to be transformed into kinetic energy during the expansion \cite{smerzi97,LDM2009}, and the bulk of the system is effectively localized. An example is shown in Fig.~\ref{fig5}, where the self-trapping is achieved by introducing an attractive interaction on an initial state that is close to the single-particle ground state of the system.  Here, we observe an interplay of noise and interaction resulting in a breakdown of the additivity and in a transient superdiffusive behavior, i.e. an expansion with $\alpha>0.5$. In absence of noise, the central part of the distribution, where $E_{int}$ is large, cannot expand, and therefore the rms width $\sigma$ increases slower than normal. Noise can break this trapping by providing the necessary energy to couple the interaction-shifted states to the lattice band. This causes the fast expansion at short times, which is even faster than normal diffusion. We note that related superdiffusion effects have been predicted in linear systems with varying energy \cite{hanggi}. An analogous effect might arise when the noise is able to release the large kinetic energy associated to the self-trapped part of the system, but further studies are needed to clarify this mechanism.\\

\begin{large}\textbf{Discussion}\end{large}\\

\begin{figure}[t]
\includegraphics[width=1.\columnwidth,clip]{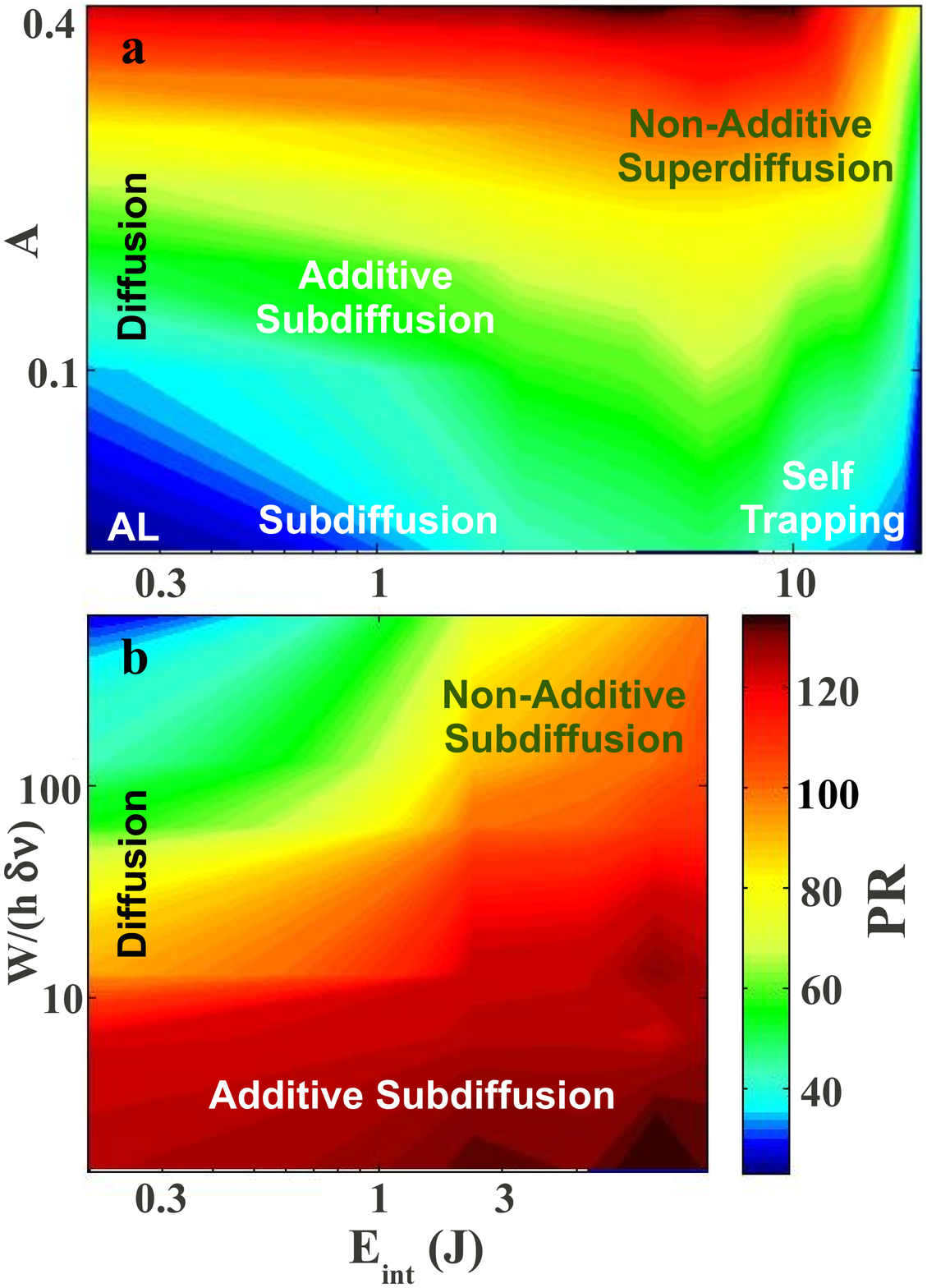}
\caption{\textbf{Generalized diffusion regimes.} The color scale represents the number of significantly occupied sites after $10$s of our numerically simulated expansion, for $\Delta=2.5J$, $J=180$ Hz and a Gaussian wavepacket initially distributed over $\approx 13$ lattice sites. a) Dependence on $A$ and $E_{int}$ for a broadband noise ($\delta \nu \approx W/h$). b) Dependence on $E_{int}$ and $\delta \nu$, for a fixed noise strength $A=0.4$.} \label{fig6}
\end{figure}

An effective overview of our extensive investigation can be achieved by studying the general behavior of the expansion for varying $A$, $E_{int}$ and $\delta\nu$ in numerical simulations. The quantity that is convenient to study is the so called participation ratio, PR=$(\sum_i n_i^2)^{-1}$ (see Methods for details), which measures the number of significantly occupied lattice sites. The specific diagram in Fig.~\ref{fig6}, which shows the calculated PR after $10$s of expansion for a particular Gaussian initial distribution, can be used to summarize general properties investigated with our system.

For a broadband noise (Fig.~\ref{fig6}a, $\delta\nu\approx W/h$), one finds again that Anderson localization is suppressed by either noise or interaction, with both leading to an expansion of the system. We remark that the underlaying mechanisms for these two expansions are quite different. A weak interaction produces a coherent coupling of single-particle localized states; the resulting transport can be modeled as a coherent hopping between such states, with a characteristic subdiffusive behavior that is determined by the changing density. Noise instead drives an incoherent hopping between localized states, which result in a diffusive expansion. Once both transport mechanisms are combined, one sees an enhanced expansion, as highlighted by the increasing width in Fig.~\ref{fig6}a for finite $A$ and $E_{int}$. Our detailed time-dependent analysis indicates that this combined diffusion is well modeled by the generalized, additive diffusion model of Eq.~(\ref{eq.diff_sum}). This surprisingly simple results seems not to be limited to the region of parameters where the noise- or interaction-induced diffusions can be modeled perturbatively.

The numerical simulations reveal also a region of stronger interaction ($E_{int}\approx 4J$), where the interaction itself prevents expansion, because of the self trapping-mechanism. This is visible as a reduction of the PR for increasing $E_{int}$ in Fig.~\ref{fig6}a. Here, the addition of noise restores a diffusion, which our time-dependent analysis has shown to have an anomalous superdiffusive nature. In this regime, the model of Eq.~(\ref{eq.diff_sum}) clearly breaks down.

Finally, an interplay of noise and interaction is seen also in presence of narrowband noise, as shown in Fig.~\ref{fig6}b. Here the expansion is studied for varying inverse noise bandwidths, $W/h\delta\nu$, and $E_{int}$, for a fixed value of the noise strength ($A=0.4$). In the non interacting regime, the diffusion constant decreases for increasing $W/h\delta\nu$. The addition of the interaction recovers a faster expansion, which however cannot be described by Eq.~(\ref{eq.diff_sum}).

In conclusion, we have experimentally realized an ultracold atomic system to investigate the transport of a wave-packet in presence of controllable disorder, noise, and interaction. We have used it to characterize the noise-induced diffusion and its interplay with an interaction-induced subdiffusion. We have observed that the complex anomalous diffusion resulting by the simultaneous presence of the two processes, can be remarkably explained by a simple generalized diffusion equation in a wide range of parameters. Finally, we have numerically observed also regimes where the presence of trapping phenomena produce a more complex interplay between the noise- and interaction-induced diffusion mechanisms. It is interesting to note that a quasiperiodic lattice might allow to study also a different regime that has been extensively analyzed in theory \cite{ovchinnikov75,bouchaud92si,saul92si}, in which in absence of noise all the eigenstates are extended, while the noisy potential is disordered. This regime can be achieved simply by using a weaker disorder, below the localization threshold $(\Delta<2J)$. Preliminary numerical simulations we have performed show that in this case the noise slows the dynamics from ballistic to diffusive.

Our work has showed how the combination of an ultracold atomic system with optical lattices can be effectively employed to study quantum phenomena related to noise. Future experiments in lattices with higher dimensionality or in reduced dimensions will allow to explore the regime of strong correlations. Here there is a strong interest in understanding the behavior of open quantum systems \cite{zoller}, and how quantum phase transitions are affected by noise \cite{giamarchi}. An interesting alternative to the use of a noisy optical potential is to employ a large sample of a second atomic species with a controllable thermal distribution, which is coupled via resonant elastic scattering to the atomic system under study. This approach might allow to introduce a noise in thermal equilibrium, and to simulate phonon-related phenomena \cite{thouless,lewenstein}.\\
Furthermore, our scheme could be easily used to control the temperature of our weakly interacting bosons in presence of disorder, to investigate the many-body metal-insulator transitions at finite temperatures \cite{altshuler}.\\

\begin{large}\textbf{Methods}\end{large}\\

\textbf{Quasiperiodic potential.} The one-dimensional quasi-periodic potential is created by a primary optical lattice combined with a weaker incommensurate one:
\be
V(x)=V_1\sin^2(k_1x+\varphi_1)+V_2\gamma^2\sin^2(\gamma k_1x+\varphi_2)\,. \label{eq:quasiperiodic}
\ee
Here $k_i=2π/\lambda_i$ are the wavevectors of the lattices ($\lambda_1=1064.4$ nm and $\lambda_2=859.6$ nm) and $\gamma=\frac{\lambda_1}{\lambda_2}$ measures the commensurability of the lattices. The relevant parameters are the spacing $d$=$\lambda_1/2$, the tunneling energy $J$ (typically 150~Hz) of the primary lattice, and the disorder strength $\Delta$, which scales linearly with $V_2$ \cite{Modugno09}. Non-interacting particles in the fundamental band of this lattice are described by the Aubry-Andr\'e model \cite{Aubry} which shows a metal-insulator transition for a finite value of the disorder $\Delta=2J$ \cite{Modugno09}.\\

\textbf{Interaction energy.} The scattering length $a$ is changed by means of a broad Feshbach resonance to values ranging from $a \approx 0.1 a_0$ to about $a=300 a_0$ \cite{Roati07}. We can define a mean interaction energy per atom $E_{int}=2\pi\hbar^2 a \bar{n}/m $, where $\bar{n}$ is the mean on-site density. The atomic sample is radially trapped with a frequency $\omega_r=2\pi\times50$~Hz. The presence of the radial degrees of freedom limits $E_{int}$ to values that are not much larger than the kinetic energy $J$ or the disorder energy $\Delta$.\\

\begin{figure}[t]
\includegraphics[width=1\columnwidth]{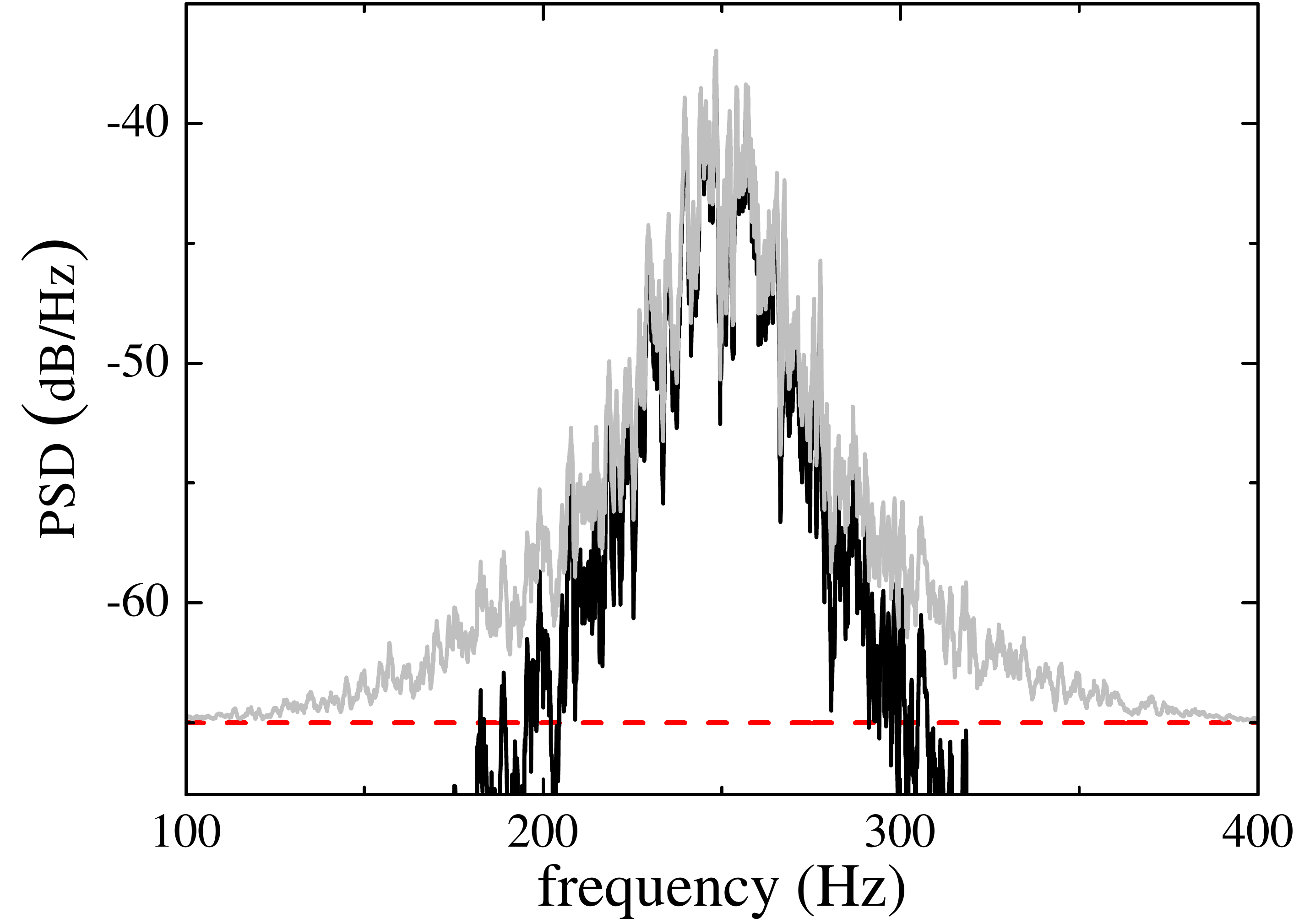}
\caption{\textbf{Noise spectrum.} Typical noise spectral density for $T=5$ ms and $A=1$ ((black-solid line). By including the broadband background noise (red-dashed line), due to acoustic noise on the lattices' mirrors, we estimate the effective experimental spectrum (grey-solid line).} \label{fig7}
\end{figure}

\textbf{Noise implementation.} In order to have a controllable amount of noise, we introduce a time dependence on the amplitude of the secondary lattice potential, which produces a time-variation of the on-site energies. In the experiment, the noise must be broadband enough to couple to as many states as possible within the lattice bandwidth ($W/ \hbar \approx(2\Delta+4J)/\hbar$) but at the same time one must avoid excitation of the radial modes ($\omega_r=2\pi\times50$~Hz) as well of the second band of the lattice ($\Delta E_{gap}/h\approx$3-5~kHz). This is achieved with a sinusoidal amplitude modulation, i.e.
\be
V_2(t)=V_0(1+A\sin(\omega_m t+\phi_m))\,,
\ee
with the frequency $\omega_{m}$ that is randomly varied in a finite interval $\omega_m\in[\omega_0-\delta\omega, \ \omega_0-\delta\omega]$ with a time step $T$, while the phase $\phi_m$ is adjusted in order to preserve the continuity of the modulation and the sign of its first derivative, to avoid frequency components outside the chosen band. The values used in the experiment, whose typical spectrum is shown in Fig.~\ref{fig7}, are: $\omega_0/2\pi=250$~Hz, $\delta\omega_0/2\pi=50$~Hz, $T=5$ ms. At long times, $t\gg T$ the resulting noise spectrum decays exponentially outside the band $[\omega_0-\delta\omega, \ \omega_0+\delta\omega]$, on a frequency scale $\approx 1/T$. Note that at short times, $t\approx T$, this spectrum contains only few components, which might result in an inefficient excitation of the localized states to provide diffusion \cite{yamadasi}. In the experiment however we do not see such effect. We believe that this is due to an additional broadband background noise in the relative phase of the two lattices, which is due to fluctuations of the position of the retroreflecting mirror at acoustic frequencies. We measured this phase noise via a Michelson interferometer scheme, and we estimate it to be equivalent to a broadband amplitude noise on the secondary lattice, about $25$ dB below the main noise when $A=1$. The estimated value is consistent with the small diffusion we have for $A=0$ and with the diffusion we observe for $A>0$ and a single frequency modulation, i.e. $\delta \omega=0$.

In the numerical simulations we find, however, a reduced expansion until $t\gg T$. For this reason, we have numerically tested also a second noise scheme corresponding to a larger width of the power spectrum already at short times and is able to better reproduce the experimental combination of the external frequency noise and the uncontrollable phase one. In this scheme the frequency $\omega_m$ is kept fixed, while the phase $\phi_m$ is randomly varied every time step $T$. Here, the power spectrum is essentially a sinc function with width $\delta\omega=2\pi/T$ as soon as $t>T$.

We have checked with extensive simulations that both types of noise lead to normal diffusion at long times. In the second case, we observe that the diffusion coefficient $D$ grows linearly with the bandwidth of the noise as long as such bandwidth is smaller than the lattice bandwidth $W/\hbar$. Larger bandwidths, i.e. very small $T$, actually lead to a reduction of $D$, since not all the frequencies can effectively produce hopping between the localized states.\\

\textbf{The theoretical model.} The numerical simulations presented in the paper are based on a generalized version of the Aubry-Andr\'{e} model \cite{Aubry}, including a mean-field interaction term \cite{LDM2009,Modugno09}:
\bea\label{DSE_2}
i\dot{\psi}_j(t) = -(\psi_{j+1}(t)+\psi_{j-1}(t))+V_{j}|\psi_j(t)|^2+\beta|\psi_j|^2\psi_j\; \ \ \ \ \
\eea
where $V_{j}=\Delta/J\sin(2\pi\gamma j)$, $\beta$ is the interaction strength and $\psi_j(t)$ are the coefficients of the wave function in the Wannier basis, normalized in such a way that their squared modulus corresponds to the atom density on the j-th site of the lattice. All energies are measured in units of the next-neighbour tunnelling energy $J$, while the natural units for time are $\hbar/J$. Note that this model can describe only the lowest energy band of the real system. Excited bands are however not populated in the experiment. The interaction parameter $\beta$ in the model can be connected to the mean interaction energy per particle in the real system, which is defined as
\be
E_{int}=\frac{2\pi\hbar^2}{m} a \frac{\int{n({\bf r})^2 d^3r}}{\int{n({\bf r}) d^3r}}\,.
\ee
Here $n({\bf r})$ is the mean on-site density distribution, i.e. the solution of the interacting Gross-Pitaevskii problem in a single well, normalized to an atom number $N/\bar{n}_s$, where $\bar{n}_s$ is the mean number of sites occupied by the atomic distribution. The relation between $E_{int}$ and the interaction strength in the model is approximately $E_{int}\approx 2 J\beta/\bar{n}_s$ \cite{Modugno09,LDM2009}.\\

To quantify the localization we consider two quantities: the width of the wave packet measured as the square root of the second moment of the spatial distribution $|\psi_j(t)|^2$,
\be
\sigma (t) = \sqrt{m_2(t)} = \Big{[}\sum_j(j-\langle j \rangle)^2 \left|\psi_j(t) \right|^2\Big{]}^{1/2}\;,
\ee
and the participation ratio PR,
\be
PR(t)=\frac{1}{\sum_j {\left|\psi_j(t)\right|^4}}\;,
\ee
measuring the number of significantly occupied lattice sites. The quantity $\langle j \rangle$ represents the average over the spatial distribution, defined as $\langle j \rangle = \sum_j j\left|\psi_j \right|^2$.\\

\textbf{Perturbative modelling the noise-induced diffusion.} In this section we consider a perturbative approach to describe the diffusion induced by noise.
The dynamics of the disordered system in absence of temporal noise is described by Eq. (\ref{DSE_2}). In the regime of small noise amplitude ($A \ll 1$), we can assume that the only effect of noise is to induce hopping between different localized states of the imperturbed system. The coupling is driven by the frequency component of the noise spectrum that is resonant with the energy difference between the states. In the regime of $\xi \approx d$ investigated in this work, we can restrict our analysis just to the coupling between neighbouring states. The hopping rate depends on the noise strength and on the overlap between the noise spectrum $s(\nu)$ and the imperturbed system energy distribution $P(\nu)$:
\be
\Gamma = \frac{A^{2}J}{\hbar} \eta \left(\frac{\Delta}{J}\right)^{2}|\langle i|\sin (2 \pi \gamma j)|f\rangle|^{2}\; ,
\ee
where $\eta=\frac{1}{A^2}\int s(\nu) P(\nu) d \nu$ is $\lesssim 1$.
To get an analytic prediction of the effect of the noise, we approximate the energy difference between states as a flat spectrum with the same width as the lattice band, $W/h \approx (2\Delta+4J)/h$.
\begin{figure}[t]
\includegraphics[width=1.\columnwidth]{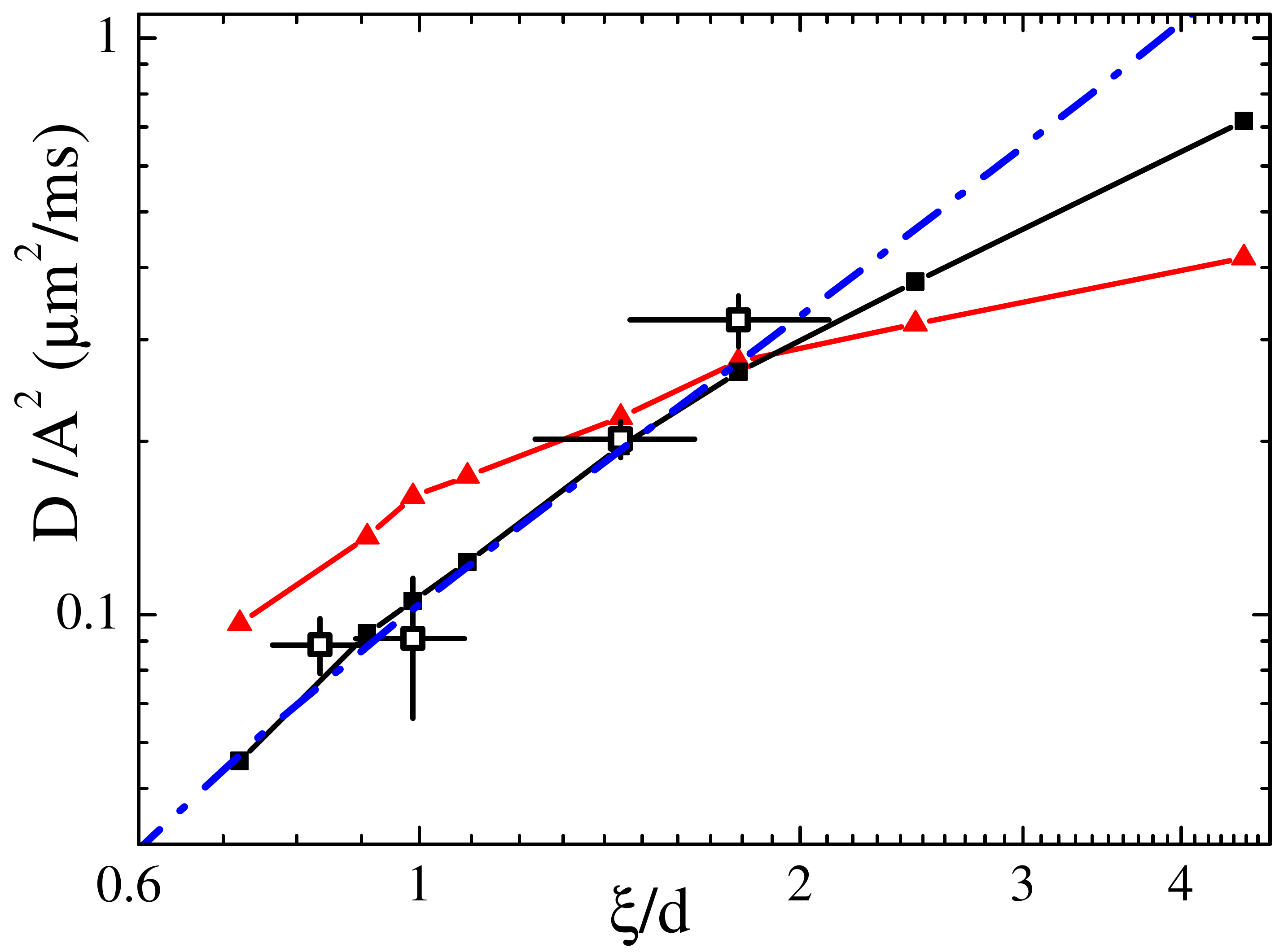}
\caption{\textbf{Noise-induced normal diffusion: $\xi$ dependence of the diffusion coefficients.} Normalized diffusion coefficient $D/A^2$ vs. the localization length $\xi$, for the regime of small noise strength $A$ in both experiment (open squares) and numerical simulations (filled squares), and for $A=1$ in simulations (triangles). The dash-dotted line shows the prediction with the perturbative approach. For numerical data, solid lines are just guides to the eye.}  \label{fig8}
\end{figure}
We approximate the overlap $|\langle i|\sin (2 \pi \gamma j)|f\rangle |^{2}$ with $|\langle i|f\rangle |^{2}$, where the states $|i\rangle $ and $|f\rangle $ are assumed to be exponentially localized over a distance $\xi$. It is also reasonable to consider $\Delta/J \sim 2 e^{d/\xi}$. Finally, when the noise spectrum is a sinc function with $T$ equal to the inverse of the bandwidth $W/h$, we obtain an analytical estimation of the diffusion coefficient
\be \label{eq.DiffCoeff}
D = \xi^{2}\Gamma \approx \frac{A^{2}J}{3\hbar} \frac{(\xi+d)^{2}}{1+e^{d/\xi}}\; .
\ee
This expression is evaluated for the optimal case of $T=h/W$, which determines the specific numerical prefactor $1/3$. In the general case we find a more complex dependence of $D$ on $\xi$, and a coefficient $D$ that decreases with the width of the noise spectrum $\delta \nu = 1/T$, i.e. the overlap between $s(\nu)$ and $P(\nu)$ decreases.
As shown in Fig.~\ref{fig8}, the diffusion coefficient obtained with this heuristic model does almost perfectly capture the evolution of $D$ with both $A$ and $\xi$. Nevertheless, we numerically observe that the perturbation approach does not hold for large values of $A$ and $\xi$, i.e. $\xi \geqslant 2d$. In these regimes, in fact, the $\xi$ dependence of $D$ becomes weaker. The good agreement between the perturbative model and the observed evolution of $D$
confirms the picture that this diffusion is induced by the hopping between localized states, in analogy with other disordered noisy systems, such as the kicked rotor with noise \cite{ott}.\\

We can estimate that the perturbative approach fails when the energy associated to the perturbation rate becomes comparable with the mean separation energy between states in a localization volume, i.e. $\hbar \Gamma \approx \Delta \frac{d}{3 \xi}$. This corresponds to a critical noise amplitude
\be \label{eq.Ac}
A_{c} \approx \Big{(}2 e^{d/\xi} \frac{1+e^{d/\xi}}{(1+\frac{d}{\xi})^{2}}\frac{d}{3 \xi}\Big{)}^{1/2}\; .
\ee
As shown in Fig.~\ref{fig2}, $A_c$ and the relative $D$, calculated respectively using Eq. (\ref{eq.Ac}) and Eq. (\ref{eq.DiffCoeff}), correctly capture the order of magnitude for the transition from the perturbative regime into a new regime in which the $\xi$ dependence of $D$ is weaker. On top of that, the fact that $A_c$ increases when $\xi$ decreases, i.e when the disorder is stronger, is also reproduced by our simulations.

There is a further regime of strong disorder that can in principle be conceived, in which the localization length is smaller than the lattice spacing and $d$ becomes therefore the relevant length-scale for the diffusion \cite{ott}. This is so far not accessible in the experiment, because the resulting diffusion is too slow to be accurately characterized.\\

\begin{large}\textbf{Supplementary Information}\end{large}\\

\begin{figure}[t]
\includegraphics[width=0.9\columnwidth]{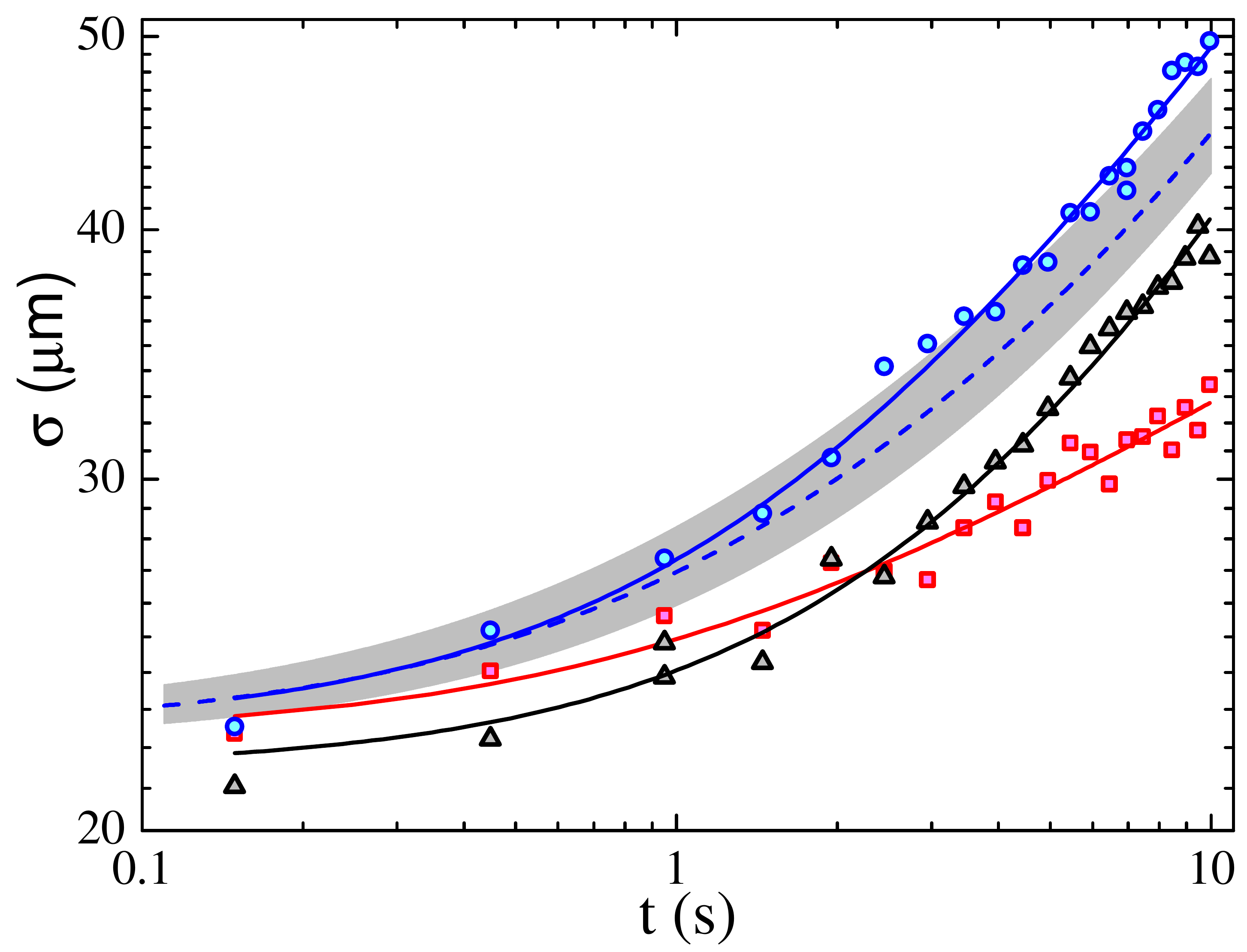}
\caption{\textbf{Noise and interaction additive anomalous diffusion.} Measured time evolution of the width for noise alone (triangles), interaction alone (squares), or both (circles), for $\Delta/J=4$ and $J = 150$~Hz. The noise amplitude is $A=0.75$, while the interaction strength is $E_{int} \sim 0.8 J$. The experimental data are fitted with Eq.~(\ref{eq.fit}) (solid lines). The dashed line is the numerical solution of Eq.~(\ref{eq.diff_sum}), with the confidence interval shown as a grey area.}  \label{SI-Fig1}
\end{figure}

\textbf{Noise- and interaction-induced delocalization.}
The conjecture summarized by Eq. (\ref{eq.diff_sum}), i.e. that noise- and interaction-induced delocalization mechanisms cooperate, is found to be valid over a broad range of parameters. In the experiment we have explored a broad range of values of both $A$ and $E_{int}$, including attractive interactions. For example, in Fig.~\ref{SI-Fig1} we show a measurement for an interaction strength similar to the one in Fig.~\ref{fig3}, but with a larger noise amplitude. The behaviour we observe is always the same: the additivity conjecture holds, except for a slightly faster expansion in presence of both noise and interaction. 
\begin{figure}[h]
\includegraphics[width=0.9\columnwidth]{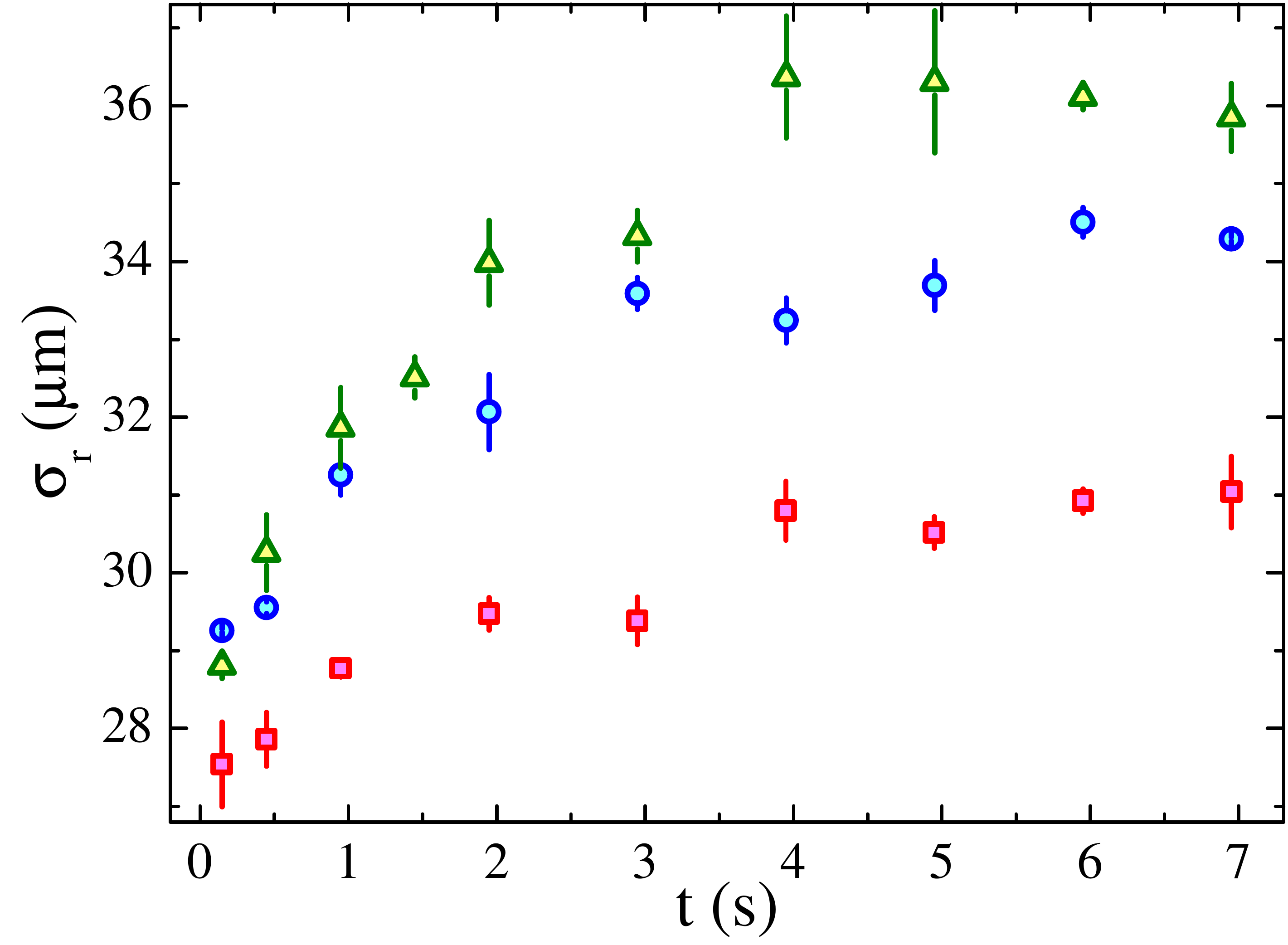}
\caption{\textbf{Increasing kinetic energy in presence of noise.} Measured time evolution of the radial width $\sigma_r$ for $E_{int} \sim 0.8 J$. Each data set corresponds to a different value of the noise amplitude: interaction alone, i.e. $A=0$ (squares), $A=0.6$ (circles) and $A=0.8$ (triangles). The parameters for the static disorder are $\Delta/J=4$ and $J = 150$~Hz.} \label{SI-Fig2}
\end{figure}

This effect is due to an increased kinetic energy in presence of noise, which is coupled to the radial degrees of freedom in presence of the interaction \cite{lucioni10}. In the experiment we can clearly detect the noise-introduced energy by probing the in situ radial size of the cloud $\sigma_r$ (related to the heating). We expect that the average kinetic energy scales approximately as $\sigma_r^2$. Fig.~\ref{SI-Fig2} shows that, in the presence of both noise and interaction, the heating of the sample is enhanced compared to the case of interaction alone. For $A=0$ just a small heating is present, presumably due to the background phase noise. This is due to the fact that the time-dependent potential continuously injects energy into the system.\\

\begin{flushleft}
\begin{large}\textbf{References}\end{large}
\end{flushleft}

\
\newline
\begin{large}\textbf{Acknowledgments}\end{large}\\

We thank Lorenzo Gori for valuable contributions in the experiment, and Shmuel Fishman and Michele Modugno for fruitful discussions. This work was supported by the ERC projects QUPOL and DISQUA, the EU projects Q-ESSENCE, AQUTE, the EU Marie-Curie Programme, MIUR-FIRB (Project No. RBFR10M3SB), and the Alexander von Humboldt Foundation. B.D. acknowledges support by the Carl-Zeiss-Stiftung.\\
\
\newline


\begin{thebibliography}{99}


\bibitem{kramer} Kramer, B., MacKinnon, A. Localization: theory and
experiment. \textit{Rep. Prog. Phys.} \textbf{56,} 1469 (1993).

\bibitem{spin glasses} Bouchard, J.-P., Dean, Aging on Parisi's tree.
\textit{Journal de Physique I}  \textbf{5,} 265-286 (1995).

\bibitem{motors}
H\"{a}nggi, P., Marchesoni, F. Artificial brownian motors: controlling
transport on the nanoscale. \textit{Rev. Mod. Phys.} \textbf{81,}
387-442 (2009).

\bibitem{PHE2004}
Plenio, M.B., Hartley, J., Eisert, J. Dynamics and manipulation of
entanglement in coupled harmonic systems with many degrees of freedom.
\textit{New J. Phys.} \textbf{6,} 36 (2004).

\bibitem{CHP2010}
Caruso, F., Huelga, F., Plenio, M.B. Noise-enhanced classical and
quantum capacities in communication networks. \textit{Phys. Rev.
Lett.} \textbf{105,} 190501 (2010).

\bibitem{MRLA2008}
Mohseni, M., Rebentrost, P., Lloyd, S., Aspuru-Guzik, A.
Environment-assisted quantum walks in photosynthetic energy transfer.
\textit{J. Chem. Phys.} {\bf 129,} 174106 (2008).

\bibitem{PH2008}
Plenio, M.B., Huelga, S.F. Dephasing-assisted transport: quantum
networks and biomolecules. \textit{New J. Phys.} {\bf 10,} 113019
(2008).

\bibitem{CLFJ2008}
Olaya-Castro, A., Lee, C.F., Fassioli Olsen, F., Johnson, F.N.
Efficiency of energy transfer in a light-harvesting system under
quantum coherence. \textit{Phys. Rev. B} \textbf{78,} 085115 (2008).

\bibitem{CCDHP2009a}
Caruso, F., Chin, A.W., Datta, A., Huelga, S.F., Plenio, M.B.
Entanglement and entangling power of the dynamics in light-harvesting
complexes. \textit{Phys. Rev. A} \textbf{81,} 062346 (2010).

\bibitem{CCDHP2009b}
Caruso, F., Chin, A.W., Datta, A., Huelga, S.F., Plenio, M.B.
Highly efficient energy excitation transfer in light-harvesting complexes: The fundamental role of noise-assisted transport.
\textit{J. Chem. Phys.} \textbf{131,} 105106 (2009).

\bibitem{White2010}
Broome, M.A., Fedrizzi, A., Lanyon, B.P., Kassa, I., Aspuru-Guzik, A.,
White, A.G. Discrete single-photon quantum walks with tunable
decoherence. \textit{Phys. Rev. Lett.} \textbf{104,} 153602 (2010).

\bibitem{Schreiber2010}
Schreiber, A., Cassemiro, K.N., Potocek, V., G\'{a}bris, A., Mosley,
P.J., Andersson, E., Jex, I., Silberhorn, Ch. Photons walking the
line: a quantum walk with adjustable coin operations.
\textit{Phys. Rev. Lett.} \textbf{104,} 050502 (2010).

\bibitem{Schreiber2012}
Schreiber, A., G\'{a}bris, A., Rohde, P.P., Laiho, K.,
\u{S}tefan\'{a}k, M., Potocek, V., Hamilton, C., Jex, I., Silberhorn,
Ch. A 2D quantum walk simulation of two-particle dynamics.
\textit{Science} \textbf{336,} 55 (2012).

\bibitem{segev}
Krivolapov, Y., Levi, L., Fishman, S., Segev, M., Wilkinson,
M. Super-diffusion in optical realizations of Anderson localization.
\textit{New J. Phys.} \textbf{14}, 043047 (2012).

\bibitem{Georges} Bouchard, J.-P., Georges, A. Anomalous diffusion in
disordered media: Statistical mechanisms, models and physical
applications. \textit{Physics Reports} \textbf{195,} 127-293 (1990).

\bibitem{ovchinnikov75}
Ovchinnikov, A.A., Erikhman, N.S. \textit{JETP} \textbf{40,} 733 (1975).

\bibitem{Madhukar77}
Madhukar, A., Post, W. Exact solution for the diffusion of a particle
in a medium with site diagonal and off-diagonal dynamic disorder.
\textit{Phys. Rev. Lett.} \textbf{39,} 1424 (1977).

\bibitem{ott}
Ott, E., Antonsen, T.M., Hanson, J.D. Effect of noise on
time-dependent quantum chaos. \textit{Phys. Rev. Lett.} \textbf{53,}
2187 (1984).


\bibitem{shepelyansky}
Fishman, S., Shepelyansky, D.L. Manifestation of localization in noise
induced ionization and dissociation. \textit{Europhys. Lett.}
\textbf{16,} 643-648 (1991).

\bibitem{cohen}
Cohen, D. Localization, dynamical correlations, and the effect of
colored noise on coherence. \textit{Phys. Rev. Lett.} \textbf{67,}
1945 (1991).

\bibitem{bayfield}
Bayfield, J.E. Near-classical noise enhancement of microwave
ionization of Rydberg atoms. \textit{Chaos} \textbf{1,} 110 (1991).

\bibitem{blumel}
Bl\"umel, R., Graham, R., Sirko, L., Smilansky, U., Walther, H.,
Yamada, K. Microwave excitation of Rydberg atoms in the presence of
noise. \textit{Phys. Rev. Lett.} \textbf{62,} 341 (1989).

\bibitem{walther}
Arndt, M., Buchleitner, A., Mantegna, R.N., Walther, H. Experimental
study of quantum and classical limits in microwave ionization of
rubidium Rydberg atoms. \textit{Phys. Rev. Lett.} \textbf{67,} 2435
(1991).

\bibitem{Steck}
Steck, D.A., Milner, V., Oskay, W.H., Raizen, M.G. Quantitative study
of amplitude noise effects on dynamical localization. \textit{Phys.
Rev. E} \textbf{62,} 3461 (2000).

\bibitem{photons}
Lahini, Y., Avidan, A., Pozzi, F., Sorel, M., Morandotti, R., Christodoulides, D. N.,  Silberberg, Y. Anderson Localization and Nonlinearity in One-Dimensional Disordered Photonic Lattices. \textit{Phys. Rev. Lett.} \textbf{100,} 013906 (2008).

\bibitem{Shepe93}
Shepelyansky, D.L. Delocalization of quantum chaos by weak
nonlinearity. \textit{Phys. Rev. Lett.} \textbf{70,} 1787 (1993).

\bibitem{Kopidakis08}
Kopidakis, G., Komineas, S., Flach, S., Aubry, S. Absence of Wave
Packet Diffusion in Disordered Nonlinear Systems. \textit{Phys. Rev.
Lett.} \textbf{100,} 084103 (2008).

\bibitem{Pikovsky08}
Pikovsky, A.S., Shepelyansky, D.L. Destruction of Anderson
localization by a weak nonlinearity. \textit{Phys. Rev. Lett.}
\textbf{100,} 094101 (2008).

\bibitem{Flach09}
Flach, S., Krimer, D.O., Skokos, C. Universal spreading of wave
packets in disordered nonlinear systems. \textit{Phys. Rev. Lett.}
\textbf{102,} 024101 (2009)

\bibitem{LDM2009}
Larcher, M., Dalfovo, F., Modugno, M. Effects of interaction on the
diffusion of atomic matter waves in one-dimensional quasiperiodic
potentials. \textit{Phys. Rev. A} \textbf{80,} 053606 (2009).

\bibitem{flach}
Flach, S. Spreading of waves in nonlinear disordered media.
\textit{Chem. Phys.} \textbf{375,} 548-556 (2010).

\bibitem{kolovsky}
Kolovsky. A.R., G\'omez, E.A., Korsch, H.J. Bose-Einstein condensates
on tilted lattices: coherent, chaotic, and subdiffusive dynamics.
\textit{Phys. Rev. A} \textbf{81,} 025603 (2010).

\bibitem{wellens}
Wellens, T., Gr\'emaud, B. Nonlinear coherent transport of waves in
disordered media. \textit{Phys. Rev. Lett.} \textbf{100,} 033902
(2008).

\bibitem{finkelstein}
Schwiete G., M. Finkelstein, A.M. Nonlinear wave-packet dynamics in a
disordered medium. \textit{Phys. Rev. Lett.} \textbf{104,} 103904
(2010).

\bibitem{cherroret}
Cherroret, N., Wellens, T. Fokker-Planck equation for transport of
wave packets in nonlinear disordered media. \textit{Phys. Rev. E}
\textbf{84,} 021114 (2011).

\bibitem{deissler2010} Deissler, B., Zaccanti, M., Roati, G.,
D'Errico, C., Fattori, M., Modugno, M., Modugno, G., Inguscio, M.
\textit{et al.} Delocalization of a disordered bosonic system by
repulsive interactions. \textit{Nature Physics} \textbf{6}, 354-358
(2010).

\bibitem{lucioni10}
Lucioni, E., Deissler, B., Tanzi, L., Roati, G., Zaccanti, M.,
Modugno, M., Larcher, M., Dalfovo, F., Inguscio, M., Modugno, G.
Observation of Subdiffusion in a Disordered Interacting System.
\textit{Phys. Rev. Lett.} \textbf{106,} 230403 (2011).

\bibitem{smerzi97}
Smerzi, A., Fantoni, S., Giovanazzi, S., Shenoy S.R.
Quantum Coherent Atomic Tunneling between Two Trapped Bose-Einstein Condensates.
\textit{Phys. Rev. Lett.} \textbf{79,} 4950 (1997).

\bibitem{mott68}
Mott, N. F.
Metal-Insulator Transition.
\textit{Rev. Mod. Phys.} \textbf{40,} 677 (1968).

\bibitem{giamarchi87}
Giamarchi, T., Schulz, H. J.
Localization and Interaction in One-Dimensional Quantum Fluids.
\textit{Europhys. Lett.} \textbf{3,} 1287  (1987).

\bibitem{fisher89}
Fisher, M. P. A., Grinstein, G., Fisher, D. S.
Boson localization and the superfluid-insulator transition.
\textit{Phys. Rev. B} \textbf{40,} 546 (1989).

\bibitem{roati2008} Roati, G., D'Errico, C., Fallani, L., Fattori, M.,
Fort, C., Zaccanti, M., Modugno, G., Modugno, M., Inguscio M. Anderson
localization of a non-interacting Bose-Einstein condensate.
\textit{Nature} \textbf{453,} 895-898 (2008).

\bibitem{Aubry}
Aubry, S., Andr\`{e}, G. Analyticity breaking and Anderson localization
in incommensurate lattices. \textit{Ann. Israel Phys. Soc.}
\textbf{3,} 133 (1980).

\bibitem{Roati07}
Roati, G., Zaccanti, M., D'Errico, C., Catani, J., Modugno, M.,
Simoni, A., Inguscio, M., Modugno, G. $^{39}$K Bose-Einstein
condensate with tunable interactions. \textit{Phys. Rev. Lett.}
\textbf{99,} 010403 (2007).

\bibitem{brow}
Aranovich, G. L., Donohue, M. D.
Diffusion Equation for Interacting Particles.
\textit{J. Phys. Chem. B} \textbf{109,} 16062 (2005).

\bibitem{Modugno09}
Modugno, M. Exponential localization in one-dimensional quasi-periodic
optical lattices. \textit{New J. Phys.} \textbf{11,} 033023 (2009).

\bibitem{larcher}
Larcher, M., Laptyeva, T.V., Bodyfelt, J.D., Dalfovo, F., Modugno, M.,
Flach, S. Subdiffusion of nonlinear waves in quasiperiodic potentials.
\textit{New J. Phys.} \textbf{14,} 103036 (2012).

\bibitem{hanggi}
Siegle, P., Goychuk, I., H\"{a}nggi, P. Origin of hyperdiffusion in
generalized brownian motion. \textit{Phys. Rev. Lett.}  \textbf{105,}
100602 (2010).

\bibitem{bouchaud92si}
Bouchaud, J.P., Toutati, D., Sornette, D. Waves in a rapidly varying
random potential: a numerical study. \textit{Phys. Rev. Lett.}
\textbf{68,} 1787 (1992).

\bibitem{saul92si}
Saul, L., Kardar, M., Read, N. Directed waves in random media.
\textit{Phys. Rev. A} \textbf{45,} 8859 (1992).

\bibitem{zoller} Diehl, S., Micheli, A., Kantian, A., Kraus, B., Buchler, H. P., Zoller, P.
Quantum states and phases in driven open quantum systems with cold atoms.
\textit{Nature Physics} \textbf{4,} 878(2008).

\bibitem{giamarchi} Dalla Torre, E. G., Demler, E., Giamarchi, T., Altman, E.
Quantum critical states and phase transitions in the presence of non-equilibrium noise.
\textit{Nature Physics} \textbf{6,} 806 (2010).

\bibitem{thouless}
Thouless, D.J.
Electrons in disordered systems and the theory of localization.
\textit{Physics Reports} \textbf{13,} 93 (1974).

\bibitem{lewenstein}
Lewenstein, M., Sanpera, A., Ahufinger, V., Damski, B., Sen(De), A.,  Sen U.
Ultracold atomic gases in optical lattices: mimicking condensed matter physics and beyond.
\textit{Adv. Phys.} \textbf{56,} 243 (2007).

\bibitem{altshuler}
Aleiner, I. L., Altshuler, B. L., Shlyapnikov, G. V.
A finite-temperature phase transition for disordered weakly interacting bosons in one dimension.
\textit{Nature Physics} \textbf{6,} 900 (2010).


\bibitem{yamadasi}
Yamada, H., Ikeda, K.S. Dynamical delocalization in one-dimensional
disordered systems with oscillatory perturbation. \textit{Phys. Rev.
E} \textbf{59,} 5214 (1999).

\end{thebibliography}
\end{document}